\documentclass[twocolumn,showpacs]{revtex4}
\usepackage[latin1]{inputenc}
\usepackage[T1]{fontenc}
\usepackage{graphicx}
\usepackage[dvips]{epsfig}
\usepackage{amsmath}
\usepackage{amssymb}
\usepackage{color}
\usepackage{dcolumn}
\usepackage{slashbox}
\usepackage{braket}

\begin{document}

\title{Pair production beyond the Schwinger formula in time-dependent electric fields}
\author{F. Hebenstreit, R. Alkofer}
\affiliation{Institut f\"ur Physik, Karl-Franzens Universit\"at Graz, A-8010 Graz, Austria}
\author{H. Gies}
\affiliation{Theoretisch-Physikalisches Institut, Friedrich-Schiller Universit\"at Jena, D-07743 Jena, Germany}
\date{\today}

\begin{abstract}
We investigate electron-positron pair production in pulse-shaped electric background fields using a non-Markovian quantum kinetic equation. We identify a pulse length range for subcritical fields still in the nonperturbative regime where the number of produced pairs significantly exceeds that of a naive expectation based on the Schwinger formula. From a conceptual viewpoint, we find a remarkable quantitative agreement between the (real time) quantum kinetic approach and the (imaginary time) effective action approach.
\end{abstract}

\maketitle

\section{Introduction}

It is a long-standing prediction of quantum electrodynamics (QED) that its vacuum is unstable in the presence of strong electric fields and decays by emitting electron-positron pairs \cite{Sauter:1931, Heisenberg:1935}. The
decay of the QED vacuum -- the so-called Schwinger mechanism -- has often been analyzed within equilibrium quantum field theory, even though it is a far-from-equilibrium, time-dependent phenomenon. For resolving the dynamics of the production process, kinetic theory including a source term for electron-positron pair production provides an appropriate approach.

According to the nonperturbative Schwinger formula \cite{Schwinger:1951nm}, which is strictly valid only for spatially homogeneous and static electric fields, a sizeable production rate requires field strengths of the order
$E_\mathrm{cr}= 1.3\cdot10^{18}\,\rm{V}/\rm{m}$. The study of corrections due to spatial or temporal inhomogeneities has a long history which includes exact solutions \cite{Nikishov:1970br,Dunne:1998ni}, various semiclassical methods
\cite{Brezin:1970xf,Popov:1972,Popov:1973az,Kim:2000un,Piazza:2004sv,Dunne:2006ff}, and functional techniques \cite{Fried:2001ga} as well as Monte Carlo simulations \cite{Gies:2005bz}. Within kinetic theory, first approaches used an instantaneous phenomenologically motivated source term based on the Schwinger formula, which served as a model for both quark-antiquark pair production in chromoelectric flux tubes and electron-positron pair production even in the case of time-dependent electric fields \cite{Casher:1978wy,Kajantie:1985jh,Gatoff:1987uf,Kluger:1991ib, Cooper:1992hw}. On the other hand, for spatially homogeneous and time-dependent electric fields, a rigorous connection between kinetic theory and a mean-field approximation to QED has been established \cite{Smolyansky:1997fc, Kluger:1998bm,Schmidt:1998vi}. It was shown that the true source term for electron-positron pair production has intrinsically non-Markovian character.

In recent years, special interest has been laid on the investigation of electron-positron pair production in alternating \cite{Alkofer:2001ik,Roberts:2002py,Blaschke:2005hs} and pulse-shaped electric fields \cite{Bloch:1999eu, Bloch:1999vh,Vinnik:2001qd} within the kinetic approach including the non-Markovian source term, with the field-current feedback due to Maxwell's equation taken into account as well. Because of the rapid development of laser technology during the past years, physicists pin their hopes on x-ray free electron laser systems and optical high-intensity laser systems to experimentally observe the Schwinger mechanism within the next decades.

In this investigation, we explore the connection bet\-ween the non-Markovian and the Schwinger-like source term for a pulse-shaped electric field in order to look for a new pair-production behavior which goes beyond the instantaneous Schwinger approximation.

\section{Particle Creation in Electric Background Fields}

We consider a spatially homogeneous, time-dependent Abelian vector potential in the temporal gauge $A_\mu(t)= (0,\vec{A}(t))$, with its spatial part defining the $\hat{z}$-direction: $\vec{A}(t)=\left(0,0,A(t)\right)$. The corresponding electric background field points into the $\hat{z}$ direction as well,
\begin{equation}
   \vec{E}(t)=-\frac{\mathrm{d}\vec{A}(t)}{\mathrm{d}t}=\left(0,0,E(t)\right)\ ,
\end{equation}
whereas the corresponding magnetic background field vanishes: $\vec{B}(t)=\nabla\times\vec{A}(t)=0$. In our calculations, we choose an electric background field of the form
\begin{equation}
  \label{eqn:efield}
  E(t)=\frac{E_0}{\cosh^2\left(t/\tau\right)} \ .
\end{equation}
The field reaches a maximum magnitude of $E_0$ at $t=0$ and switches on and off exponentially near $t\approx \pm4\tau$, where the electric background field drops to a value of approximately one per mille of the maximum magnitude. The temporal-gauge vector potential $A(t)$ giving rise to this electric background field is given by
\begin{equation}
  \label{eqn:afield}
  A(t)=-E_0\tau\left[1+\tanh{(t/\tau)}\right] \ .
\end{equation}

\subsection{Quantum kinetic equation}

A key quantity in the description of the pair-production process in the electric background field is the single-particle momentum distribution function $f(\vec{k},t)$. It can be computed from a quantum Vlasov equation
with a source term for electron-positron pair production. It receives a physical meaning as distribution function of real particles as we go to asymptotic times $t\rightarrow\pm\,\infty$ when the electric background field vanishes. In our opinion, a similar interpretation at intermediate times \cite{Blaschke:2005hs} is difficult to justify. In addition to the source term, we would in general have to take collisions and the field-current feedback due to Maxwell's equation into account; however, in the field strength regime we are mainly interested in, i.e. the subcritical field strength regime with $E_0\approx0.1E_\mathrm{cr}$, these effects can be neglected \cite{Bloch:1999eu}, and we obtain
\begin{eqnarray}
  \label{eqn:qkeq}
  \frac{\mathrm{d}f^{\mathrm{nM}}(\vec{k},t)}{\mathrm{d}t}&=&\frac{1}{2}\frac{eE(t)\epsilon_\perp}{\omega^2(\vec{k},t)}\int_{-\infty}^{t}{\mathrm{d}t'\frac{eE(t')\epsilon_\perp}{\omega^2(\vec{k},t')}} \\ 
  &\times&\left[1-2f^{\mathrm{nM}}(\vec{k},t')\right]\cos\left[2\int_{t'}^{t}{\mathrm{d}\tau\,\omega(\vec{k},\tau)}\right] \nonumber \ ,
\end{eqnarray}
with $e$ the electric charge. $\vec{k}=(\vec{k}_\perp,k_3)$ denotes the canonical three-momentum vector with $p_\parallel(t)=k_3-eA(t)$ being the kinetic momentum in the $\hat{z}$ direction, $\epsilon_\perp^2=m^2+ \vec{k}_\perp^2$ is the transverse energy squared and $\omega^2(\vec{k},t)=\epsilon_\perp^2+p_\parallel(t)^2$ characterizes the total energy squared. This equation, which is valid for both helicities, has a non-Markovian character for two reasons: the factor $[1-2f(\vec{k},t)]$ arising from quantum statistics takes the time
history of the distribution function into account, whereas the cosine factor accounts for the time history of the electric background field. In order to simplify the numerical treatment of this equation, we may introduce the auxiliary quantities
\begin{eqnarray}
  u^{\mathrm{nM}}(\vec{k},t)&=&\int_{-\infty}^{t}{\mathrm{d}t'\frac{eE(t')\epsilon_\perp}{\omega^2(\vec{k},t')}} \\
  &\times&\left[1-2f^{\mathrm{nM}}(\vec{k},t')\right]\sin\left[2\int_{t'}^{t}{\mathrm{d}\tau\,\omega(\vec{k},\tau)}\right] \nonumber \ , \\
  v^{\mathrm{nM}}(\vec{k},t)&=&\int_{-\infty}^{t}{\mathrm{d}t'\frac{eE(t')\epsilon_\perp}{\omega^2(\vec{k},t')}} \\
  &\times&\left[1-2f^{\mathrm{nM}}(\vec{k},t')\right] \cos\left[2\int_{t'}^{t}{\mathrm{d}\tau\,\omega(\vec{k},\tau)}\right] \nonumber \ ,
\end{eqnarray}
and reformulate the integro-differential equation~(\ref{eqn:qkeq}) as a coupled system of first order differential equations:
\begin{eqnarray}
  \label{eqn:nm1}
  \frac{\mathrm{d}}{\mathrm{d}t}f^{\mathrm{nM}}(\vec{k},t)&=&\frac{1}{2}\frac{eE(t)\epsilon_\perp}{\omega^2(\vec{k},t)}v^{\mathrm{nM}}(\vec{k},t) \ , \\
 \label{eqn:nm2}
  \frac{\mathrm{d}}{\mathrm{d}t}u^{\mathrm{nM}}(\vec{k},t)&=&2\omega(\vec{k},t)v^{\mathrm{nM}}(\vec{k},t) \ , \\
 \label{eqn:nm3}
  \frac{\mathrm{d}}{\mathrm{d}t}v^{\mathrm{nM}}(\vec{k},t)&=&\frac{eE(t)\epsilon_\perp}{\omega^2(\vec{k},t)}\left[1-2f^{\mathrm{nM}}(\vec{k},t)\right] \\
   & &-2\omega(\vec{k},t)u^{\mathrm{nM}}(\vec{k},t) \nonumber .
\end{eqnarray}
While the non-Markovian character of Eq.~(\ref{eqn:qkeq}) due to the cosine factor is crucial for the electron-positron pair creation process, we anticipate that memory effects due to the statistical factor become only
important for very strong electric background fields of the order of $E_0\approx E_\mathrm{cr}$. Then, the Compton time $t_{\text{c}}=1/m$ is of the order of the time scale of electron-positron pair creation \cite{Labun:2008}, and a sizeable distribution localized in phase space is accumulated such that statistical effects such as Pauli blocking become relevant. However, in the field strength regime we are interested in,  $E_0\approx0.1E_\mathrm{cr}$, neglecting the statistical factor for $f(\vec{k},t)\ll1$ yields the low-density approximation:
\begin{eqnarray}
  \label{eqn:ldeq}
  \frac{\mathrm{d}}{\mathrm{d}t}f^{\mathrm{ld}}(\vec{k},t)&=&\frac{1}{2}\frac{eE(t)\epsilon_\perp}{\omega^2(\vec{k},t)}\int_{-\infty}^{t}{\mathrm{d}t'\frac{eE(t')\epsilon_\perp}{\omega^2(\vec{k},t')}} \\
  &\times&\cos\left[2\int_{t'}^{t} {\mathrm{d}\tau\,\omega(\vec{k},\tau)}\right] \nonumber \ .
\end{eqnarray}
This equation can be solved either by direct integration
\begin{eqnarray}
  f^{\mathrm{ld}}(\vec{k},t)&=&\frac{1}{2}\int_{-\infty}^{t}{\mathrm{d}t''\frac{eE(t'')\epsilon_\perp}{\omega^2(\vec{k},t'')}} \\
  &\times&\int_{-\infty}^{t''}{\mathrm{d}t'\frac{eE(t')\epsilon_\perp}{\omega^2(\vec{k},t')} \cos\left[2\int_{t'}^{t''} {\mathrm{d}\tau\,\omega(\vec{k},\tau)}\right]} \nonumber \ , 
\end{eqnarray}
or, as before, by reformulating it as a coupled system of first-order differential equations:
\begin{eqnarray}
  \label{eqn:ld1}
  \frac{\mathrm{d}}{\mathrm{d}t}f^{\mathrm{ld}}(\vec{k},t)&=&\frac{1}{2}\frac{eE(t)\epsilon_\perp}{\omega^2(\vec{k},t)}v^{\mathrm{ld}}(\vec{k},t) \ , \\
  \label{eqn:ld2}
  \frac{\mathrm{d}}{\mathrm{d}t}u^{\mathrm{ld}}(\vec{k},t)&=&2\omega(\vec{k},t)v^{\mathrm{ld}}(\vec{k},t)
  \label{eqn:ld3} \ , \\
  \frac{\mathrm{d}}{\mathrm{d}t}v^{\mathrm{ld}}(\vec{k},t)&=&\frac{eE(t)\epsilon_\perp}{\omega^2(\vec{k},t)}-2\omega(\vec{k},t)u^{\mathrm{ld}}(\vec{k},t)
   \ .
\end{eqnarray}

An important observable is given by the asymptotic particle number density $n[e^+e^-]$ which is obtained as an integral of the asymptotic distribution function $f(\vec{k},\infty)$
\begin{equation}
  \label{eqn:num}
  n[e^+e^-]=2\int{\frac{\mathrm{d}^3k}{(2\pi)^3}f(\vec{k},\infty)} \ ,
\end{equation}
with the factor of $2$ arising from spin degeneracy.

{
\subsection{Effective action approach}

We briefly summarize the main results obtained within the effective action approach to electron-positron pair production for later use. Within this approach the vacuum-to-vacuum persistence amplitude in the presence of an
external electromagnetic field $A_\mu$ is represented as
\begin{equation}
  \label{eqn:eavtv}
  \braket{0|0}^{A}=\exp{[iS_\mathrm{eff}]} \ ,
\end{equation}
with $S_\mathrm{eff}$ denoting the effective action. At weak fields, the imaginary part of the effective action is an estimate of the number of produced electron-positron pairs per unit volume \cite{Nikishov:1970br,Cohen:2008}},
\begin{equation}
  \label{eqn:eaim}
   n^{\mathrm{eff}}[e^+e^-]\simeq2\operatorname{Im}{[S_\mathrm{eff}]} \ .
\end{equation}
For the present electric field Eq.~(\ref{eqn:efield}), an exact integral representation of the imaginary part of the effective action can be found \cite{Nikishov:1970br,Dunne:1998ni}:
\begin{equation}
  \label{eqn:eanum}
  \operatorname{Im}{[S_\mathrm{eff}]} 
  =-\frac{1}{8\pi^2}\int{\mathrm{d}^3k\,\ln\left[(1-e^{-\pi\Omega_+})(1-e^{-\pi\Omega_-})\right]} \ ,
\end{equation}
with $\Omega_\pm=\tau(\alpha_++\alpha_-\pm2eE_0\tau)$ and $\alpha_\pm=\sqrt{\epsilon_\perp^2+(k_3\pm eE_0\tau)^2}$.

\subsection{Instantaneous approximation} 

Results concerning electron-positron pair production in time-dependent electric fields $E(t)$ within both the quantum kinetic approach and the effective action approach were derived long after Schwinger had found an analytic expression for the vacuum decay rate in a spatially homogeneous but static electric field $E_0$ pointing into the $\hat{z}$ direction:
\begin{equation}
  \label{eqn:snum}
  \operatorname{Im}{[S_\mathrm{eff}]}=\frac{e^2E_0^2}{8\pi^3}\sum_{n=1}^{\infty}{\frac{1}{n^2}\exp\left(-\frac{n\pi m^2}{eE_0}\right)} \ . 
\end{equation}
As an ansatz, this result was assumed to approximately hold for arbitrary time-dependent electric background field $E(t)$ as well. This served as a starting point for formulating a Vlasov equation for the single-particle
momentum distribution function including an instantaneous phenomenologically motivated source term for electron-positron pair production \cite{Casher:1978wy,Kluger:1991ib,Cooper:1992hw}:
\begin{eqnarray}
  \label{eqn:seq}
  \frac{\mathrm{d}}{\mathrm{d}t}f^\mathrm{S}(\vec{k},t)&=&-eE(t)\,\delta(k_3-eA(t)) \nonumber \\
  & \times&\ln\left[1-\exp\left(-\frac{\pi(m^2+\vec{k}_\perp^2)}{eE(t)}\right)\right] \ .
\end{eqnarray}
The instantaneous source term is chosen in such a way that Eq.~(\ref{eqn:snum}) is recovered [with $E_0$ being replaced by $E(t)$] upon integrating over the whole momentum space and by including a factor of $2$ for spin degeneracy. Of course, using the instantaneous source term is not justified in the presence of a rapidly varying electric background; deviations of the full result from this approximation for time-dependent fields $E(t)$ point to new properties of pair production beyond the Schwinger formula, which will be investigated in the next section.

\section{Numerical Results}

In our studies, we compare $n[e^+e^-]$ obtained by means of the numerical solution of both the non-Markovian equation~(\ref{eqn:qkeq}) and the low-density approximation~(\ref{eqn:ldeq}) with the exact effective action result Eq.~(\ref{eqn:eanum}) and the instantaneous approximation~(\ref{eqn:snum}), in order to look for new pair-production properties in time-dependent electric background fields $E(t)$ which are unexpected from the Schwinger formula~(\ref{eqn:snum}). We also check the validity of the low-density approximation in this field strength regime. The numerical solutions are found by first solving the coupled system of first-order differential equations~(\ref{eqn:nm1})$-$(\ref{eqn:nm3}) and (\ref{eqn:ld1})$-$(\ref{eqn:ld3}), respectively. We use an adaptive
Runge-Kutta method and then perform the momentum integral Eq.~(\ref{eqn:num}) on a finite momentum grid.

Obviously, the asymptotic particle number density within the instantaneous approximation should be directly proportional to the temporal width $\tau$ of the electric pulse; cf. Eq.~(\ref{eqn:snum}):
\begin{eqnarray}
  n^S[e^+e^-]&=&\frac{e^2E_0^2\tau}{4\pi^3}\int_{-\infty}^{\infty}{\mathrm{d}t'} \\
  &\times&\frac{1}{\cosh^4(t')}\exp\left(-\frac{\pi m^2\cosh^2(t')}{eE_0}\right). \nonumber \ 
\end{eqnarray}

\subsection{Pulse length and field strength dependence}

\begin{figure}[t]
  \includegraphics{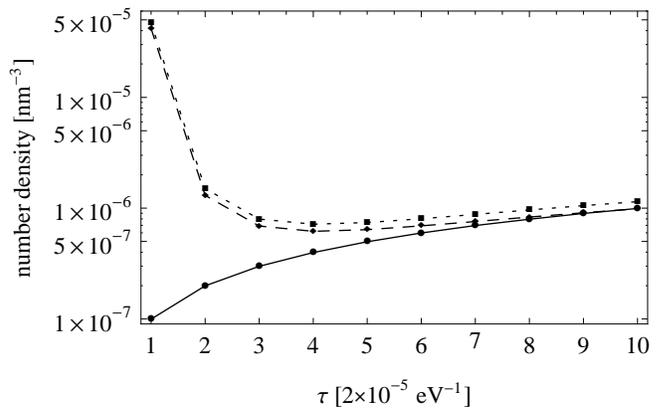}
  \caption{\label{fig:pulse} Logarithmic plot of the number density $n[e^+e^-]$ calculated for $E_0=0.1E_\mathrm{cr}$ as a function of the temporal width $\tau$ in units of $[2\cdot10^{-5}\,\rm{eV}^{-1}]$. Dashed line: solution obtained for the non-Markovian equation~(\ref{eqn:qkeq}) and within the effective action approach Eq.~(\ref{eqn:eanum}). Dotted line: solution obtained within the low-density approximation Eq.~(\ref{eqn:ldeq}). Solid line: solution obtained in the instantaneous approximation Eq.~(\ref{eqn:snum}).}
\end{figure}

\begin{table}[b]
  \caption{\label{tab:snm} $n^{\mathrm{nM}}[e^+e^-]/n^{\mathrm{S}}[e^+e^-]$ as a function of the temporal width $\tau$ in units of $[2\cdot10^{-5}\,\rm{eV}^{-1}]$ and the ratio $\mathcal{E}=E_0/E_\mathrm{cr}$. Especially in the vicinity of $\tau\approx10\cdot t_\mathrm{c}$ and $E_0=0.1E_\mathrm{cr}$ we obtain a vast discrepancy. Points with deviations greater than $10\%$ are labeled boldface.} 
\begin{ruledtabular}
\begin{tabular}{c|cccccccc}
\backslashbox{$\mathcal{E}$}{$\tau$}&1&2&3&4&5&6&7&8\\
\hline
0.1&\begin{footnotesize}\textbf{420.4}\end{footnotesize}&\begin{footnotesize}\textbf{6.568}\end{footnotesize}&\begin{footnotesize}\textbf{2.311}\end{footnotesize}&\begin{footnotesize}\textbf{1.555}\end{footnotesize}&\begin{footnotesize}\textbf{1.286}\end{footnotesize}&\begin{footnotesize}\textbf{1.159}\end{footnotesize}&\begin{footnotesize}1.087\end{footnotesize}&\begin{footnotesize}1.043\end{footnotesize}\\
0.2&\begin{footnotesize}\textbf{2.969}\end{footnotesize}&\begin{footnotesize}\textbf{1.295}\end{footnotesize}&\begin{footnotesize}1.093\end{footnotesize}&\begin{footnotesize}1.029\end{footnotesize}&\begin{footnotesize}1.001\end{footnotesize}&\begin{footnotesize}0.985\end{footnotesize}&&\\
0.3&\begin{footnotesize}\textbf{1.452}\end{footnotesize}&\begin{footnotesize}1.073\end{footnotesize}&\begin{footnotesize}1.012\end{footnotesize}&\begin{footnotesize}0.991\end{footnotesize}&\begin{footnotesize}0.982\end{footnotesize}&&&\\
0.4&\begin{footnotesize}\textbf{1.187}\end{footnotesize}&\begin{footnotesize}1.023\end{footnotesize}&&&&&&\\
0.5&\begin{footnotesize}1.096\end{footnotesize}&&&&&&&\\
0.6&\begin{footnotesize}1.056\end{footnotesize}&&&&&&&\\
0.7&\begin{footnotesize}1.035\end{footnotesize}&&&&&&&\\
0.8&\begin{footnotesize}1.022\end{footnotesize}&&&&&&&\\
0.9&\begin{footnotesize}1.014\end{footnotesize}&&&&&&&\\
1&\begin{footnotesize}1.009\end{footnotesize}&&&&&&&\\
\end{tabular}
\end{ruledtabular}
\end{table}

\begin{figure}[t]
  \includegraphics[width=7cm]{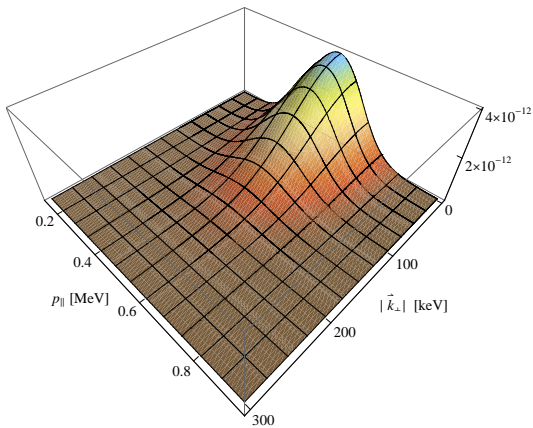}
  \caption{\label{fig:momspace} Momentum space plot of the asymptotic distribution function $f^{\mathrm{nM}}(\vec{k},\infty)$ calculated for $E_0=0.1E_\mathrm{cr}$ and $\tau=2\cdot10^{-5}\,\rm{eV}^{-1} \approx10\cdot t_\mathrm{c}$. It is an advantage of the quantum kinetic approach that the position and the width of the peak of the distribution function in momentum space can be calculated very precisely.}
\end{figure}

As a first aim, we explore the asymptotic particle number density $n[e^+e^-]$ reached in an electric background field with $E_0=0.1E_\mathrm{cr}$ as a function of the temporal width $\tau$ ranging from $2\cdot10^{-5}\,\rm{eV}^{-1}$ to $2\cdot10^{-4}\,\rm{eV}^{-1}$, cf. Fig.~\ref{fig:pulse} and Table~\ref{tab:snm}. The shortest time scale has been chosen such that the Keldysh adiabaticity parameter $\gamma=E_{\mathrm{cr}}/(E m \tau)\simeq 1$, separating the nonperturbative Schwinger regime $\gamma\ll 1$ from the perturbative multiphoton region $\gamma\gg 1$.

First we note that the results obtained by means of the non-Markovian equation~(\ref{eqn:qkeq}) and the effective action approach Eq.~(\ref{eqn:eanum}) coincide up to the third digit, with the numerical deviation arising from the finite numerical accuracy. Hence, we have shown the equivalence between the (imaginary time) effective action approach and the (real time) quantum kinetic approach. Moreover, the quantum kinetic calculation contains in addition to the particle number density $n[e^+e^-]$ useful information about the momentum space distribution $f(\vec{k},\infty)$ of the electron-positron pairs, cf. Fig.~\ref{fig:momspace}.

Second, comparing the results obtained by means of the non-Markovian equation~(\ref{eqn:qkeq}) and the instantaneous approximation Eq.~(\ref{eqn:snum}) indicates that the latter significantly underestimates the reachable particle number density at short times. For example, deviations greater than a factor of $400$ occur at extreme short time scales $\tau=2\cdot10^{-5}\,\rm{eV}^{-1} \approx 10\cdot t_\mathrm{c}$, with $t_\mathrm{c}$ corresponding to the Compton time. The Schwinger result is obtained again for temporal widths $\tau\gtrsim 70\cdot t_\mathrm{c}$. 
Our results confirm the observation that time variations of the field generically increase the pair-production yield \cite{Popov:1972,Dunne:2006ff}. 

\begin{figure}[t]
  \includegraphics{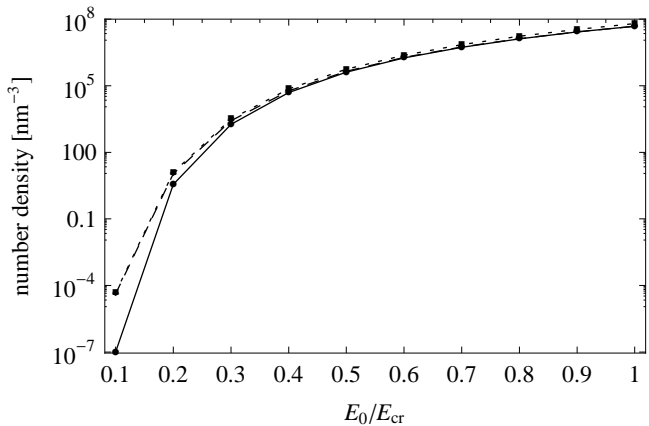}
  \caption{\label{fig:field} Logarithmic plot of the number density calculated for $\tau=2\cdot10^{-5}\,\rm{eV}^{-1} \approx 10\cdot t_\mathrm{c}$ as a function of the field strength $E_0$. Dashed line: solution obtained for the non-Markovian equation~(\ref{eqn:qkeq}) and within the effective action approach Eq.~(\ref{eqn:eanum}). Dotted line: solution obtained within the low-density approximation Eq.~(\ref{eqn:ldeq}). Solid line: solution obtained in the instantaneous approximation Eq.~(\ref{eqn:seq}).}
\end{figure}

As a second aim, we explore the asymptotic particle number density $n[e^+e^-]$ reached in an extreme short
pulse, i.~e. $\tau=2\cdot10^{-5}\,\rm{eV}^{-1}\approx 10\cdot t_\mathrm{c}$, as a function of the field strength of the electric background field with $E_0$ ranging from $0.1E_\mathrm{cr}$ to $E_\mathrm{cr}$, cf. Fig.~\ref{fig:field} and Table~\ref{tab:snm}. The numerical results indicate that the instantaneous approximation significantly
underestimates the reachable particle number density for subcritical field strengths, whereas the Schwinger result is obtained again for field strengths of the order of $E_0\gtrsim0.5E_\mathrm{cr}$ where the Keldysh parameter drops below $\gamma \simeq 0.2$.

\begin{table}[b]
  \caption{\label{tab:ldnm} $n^{\mathrm{ld}}[e^+e^-]/n^{\mathrm{nM}}[e^+e^-]$ as a function of the temporal width $\tau$ in units of $[2\cdot10^{-5}\,\rm{eV}^{-1}]$ and the ratio $\mathcal{E}=E_0/E_\mathrm{cr}$. The numerical results show a strong $\mathcal{E}$ dependence whereas the influence of the pulse length on the result is very weak.} 
\begin{ruledtabular}
\begin{tabular}{c|cccccccc}
\backslashbox{$\mathcal{E}$}{$\tau$}&1&2&3&4&5&6&7&8\\
\hline
0.1&1.127&1.150&1.156&1.158&1.159&1.159&1.160&1.160\\
0.2&1.175&1.188&1.191&1.192&1.192&1.192&&\\
0.3&1.206&1.214&1.216&1.217&1.217&&&\\
0.4&1.229&1.235&&&&&&\\
0.5&1.247&&&&&&&\\
0.6&1.263&&&&&&&\\
0.7&1.277&&&&&&&\\
0.8&1.289&&&&&&&\\
0.9&1.300&&&&&&&\\
1&1.310&&&&&&&\\
\end{tabular}
\end{ruledtabular}
\end{table}

\subsection{Non-Markovian solution vs low-density approximation}

In Table~\ref{tab:ldnm} we show the ratio $n^{\mathrm{ld}}[e^+e^-]/n^{\mathrm{nM}}[e^+e^-]$ in order to explore the validity region of the low-density approximation. For weak and static electric fields $E(t)=E_0<E_\mathrm{cr}$, the applicability of this approximation has been shown \cite{Schmidt:1998zh}. For pulse-shaped fields, the low-density approximation yields good results with deviations of the order of $10\%$ only in the field strength regime $E_0\approx0.1E_\mathrm{cr}$ but becomes unreliable when reaching higher field strengths. This deviation increases only very slowly with pulse length, whereas for increasing field strengths the deviation rises significantly. Although the quantitative results within the low-density approximation get worse for higher field strengths, the approximation does not break down completely but still gives a reasonable qualitative picture. However, for accurate predictions of the single-particle momentum distribution function $f(\vec{k},\infty)$ and the particle number density $n[e^+e^-]$ one has to take all non-Markovian effects into account. Additionally, near and beyond $E_0\approx E_\mathrm{cr}$, also the field-current feedback due to Maxwell's equation eventually needs to be taken into account. More detailed comparisons between different approximations to the quantum kinetic equation including those with backreaction can be found in \cite{Hebenstreit:2008}.

\section{Conclusions}

We have numerically investigated electron-positron pair creation in short pulse-shaped electric {background} fields by means of an existing quantum kinetic formulation. A thorough comparison is made to results obtained from an exact effective action method and a phenomenologically motivated instantaneous approximation.

One main result is the perfect agreement between the (real time) quantum kinetic approach and the (imaginary time) effective action approach. This is quite remarkable since the concepts behind these two approaches are very different: The quantum kinetic equation describes the real time nonequilibrium evolution of the quantum system whereas the effective action approach is an imaginary time equilibrium method.

Additionally, we have shown that the instantaneous approximation drastically underestimates the reachable particle number density in the field strength regime $E_0\approx0.1E_\mathrm{cr}$ when dealing with extreme short temporal widths $\tau\approx2\cdot10^{-5}\,\rm{eV}^{-1}\approx10\cdot t_\mathrm{c}$. Alongside, we have reconfirmed that the low-density approximation of the  quantum kinetic equation gives reasonable qualitative results in the subcritical field strength regime.

Our results suggest similar findings for larger pulse lengths but smaller field strengths, yielding similar Keldysh parameters $\gamma\simeq1$. Numerical studies in this regime are challenging. Given the future potential of pulse shaping with higher harmonics, and in the light of suggestions for enhancing pair production with taylored pulses \cite{Schutzhold:2008pz}, such investigations are highly relevant and clearly a worthwhile pursuit.

HG thanks the DFG for support within the Heisenberg program and the SFB-TR18.

\end{document}